\documentclass[pdflatex,sn-basic]{sn-jnl}

\usepackage{graphicx}%
\usepackage{multirow}%
\usepackage{amsmath,amssymb,amsfonts}%
\usepackage{amsthm}%
\usepackage{mathrsfs}%
\usepackage[title]{appendix}%
\usepackage{xcolor}%
\usepackage{textcomp}%
\usepackage{manyfoot}%
\usepackage{booktabs}%
\usepackage{tabularx}
\usepackage{enumitem} 
\usepackage{algorithm}%
\usepackage{algorithmicx}%
\usepackage{algpseudocode}%
\usepackage{listings}%
\usepackage{ragged2e}
\usepackage{xcolor}
\usepackage{makecell}

\theoremstyle{thmstyleone}%
\begin{document}

\title[Geocoding Literature Review]{Toward building next-generation Geocoding systems: a systematic review}


\author*[1]{\fnm{Zhengcong} \sur{Yin}}\email{yinzhengcong@tamu.edu}

\author[1]{\fnm{Daniel W.} \sur{Goldberg}}\email{daniel.goldberg@tamu.edu}

\author[1]{\fnm{Binbin} \sur{Lin}}\email{bb2020@tamu.edu}

\author[2]{\fnm{Bing} \sur{Zhou}}\email{bbz5159@psu.edu}

\author[1]{\fnm{Diya} \sur{Li}}\email{diya.li@tamu.edu}

\author[3]{\fnm{Andong} \sur{Ma}}\email{anma@msudenver.edu}

\author[4]{\fnm{Ziqian} \sur{Ming}}\email{zming@esri.com}

\author[1]{\fnm{Heng} \sur{Cai}}\email{hengcai@tamu.edu}

\author[1]{\fnm{Zhe} \sur{Zhang}}\email{zzhang@tamu.edu}

\author[5]{\fnm{Shaohua} \sur{Wang}}\email{wangshaohua@aircas.ac.cn}

\author[6]{\fnm{Shanzhen} \sur{Gao}}\email{sgao@vsu.edu}

\author[7]{\fnm{Joey Ying} \sur{Lee}}\email{humi@lazicorner.com}

\author[8]{\fnm{Xiao} \sur{Li}}\email{xiao.li@ouce.ox.ac.uk}

\author[9]{\fnm{Da} \sur{Huo}}\email{daniel.huo@utoronto.ca}

\affil[1]{\orgdiv{Department of Geography}, \orgname{Texas A\&M University}, \orgaddress{\street{797 Lamar St.}, \city{College Station}, \postcode{77840}, \state{TX}, \country{USA}}}
\affil[2]{\orgdiv{Department of Geography}, \orgname{Pennsylvania State University}, \orgaddress{\street{201 Old Main}, \city{University Park}, \postcode{16802}, \state{PA}, \country{USA}}}
\affil[3]{\orgdiv{Department of Earth and Atmospheric Sciences}, \orgname{Metropolitan State University of Denver}, \orgaddress{\street{Science Building 2014}, \city{Denver}, \postcode{80217}, \state{CO}, \country{USA}}}
\affil[4]{\orgname{Esri, Inc}, \orgaddress{\street{380 New York St.}, \city{Redlands}, \postcode{92373}, \state{CA}, \country{USA}}}
\affil[5]{\orgdiv{State Key Laboratory of Remote Sensing and Digital Earth, Aerospace Information Research Institute}, \orgname{Chinese Academy of Sciences}, \orgaddress{\street{No.9 Dengzhuang South Rd.}, \city{Beijing}, \country{China}}}
\affil[6]{\orgdiv{Department of Computer Information Systems}, \orgname{Virginia State University}, \orgaddress{\street{1 Hayden St.}, \city{Petersburg}, \postcode{23806}, \state{VA}, \country{USA}}}
\affil[7]{\orgname{LABI Education}, \orgaddress{\street{143 Keelung Rd.}, \city{Taipei}, \country{Taiwan}}}
\affil[8]{\orgdiv{Transport Studies Unit}, \orgname{University of Oxford}, \orgaddress{\street{South Parks Rd.}, \city{Oxford}, \postcode{OX1 3QY}, \country{United Kingdom}}}
\affil[9]{\orgdiv{Department of Civil \& Mineral Engineering}, \orgname{University of Toronto}, \orgaddress{\street{35 St George St.}, \city{Toronto}, \state{ON}, \country{Canada}}}


\abstract{Geocoding systems are widely used in both scientific research for spatial analysis and everyday life through location-based services. The quality of geocoded data significantly impacts subsequent processes and applications, underscoring the need for next-generation systems. In response to this demand, this review first characterizes the technical requirements for next-generation geocoding inputs and outputs. We then decompose the geocoding workflow into modular functional units and survey existing implementations. For each component, we identify methodological limitations, articulate domain-specific research questions and hypotheses, and outline evaluation strategies needed. Finally, we identify opportunities to improve next-generation geocoding systems in light of recent technological advances. We envision that this review provides a technical foundation and research agenda for advancing the design, assessment, and deployment of next-generation geocoding systems.}




\maketitle

\section{Introduction}

The geocoding process converts human-readable location descriptions into machine-readable geographic coordinates~\citep{goldberg2007text,boscoe2008science}. The earliest known example dates back over two centuries, when a British doctor employed a dot map to visualize cholera cases~\citep{mcleod2000SnowFirstGeocoding}. In recent decades, geocoding has expanded beyond academic spatial analysis into widespread everyday applications~\citep{yin2019nlpQAframework,murray2011hybrid,rose2004historical,rushton2006geocodingforHealth,li2020modeling,yin2020probabilistic,cheng2024assessing}. This expansion has been driven by improved match rates and spatial accuracy resulting from better reference datasets, enhanced address models, and diverse data sources~\citep{amelunxen2009approach,vieira2008accuracy,zandbergen2011influence}, along with rapid technological advances in web services and GPS-enabled mobile devices. Consequently, geocoding services have become readily accessible for daily use, and spatial analyses increasingly depend upon accurate geocoded data~\citep{rushton2006geocodingforHealth,murray2011hybrid}.

Significant research has examined geocoding systems from both methodological and applied perspectives. Methodological studies have focused on key components of the geocoding workflow, including address parsing~\citep{Christen2005hmmAddressParserInputErrors,mokhtari2019BILSTMaddressParser,li2018hybridChineseAddressSegmentions,yin2023chatgpt}, feature matching~\citep{Ranzijn2013AddressMatchingTechniques,lin2020W2VforGeocoding,Goldberg2011ImprovingMatchRateVaryingBlock,murray2011hybridGeocodesNearbyHouseNumber}, interpolation~\citep{goldberg2008geocodingbest,Cayo2003,Bakshi:2004:EOS:1032222.1032251}, and candidate ranking~\citep{goldberg2010candidateselectioncriteria,knoblock2017ElasticSpatialExampleGeocoding}. In parallel, comprehensive evaluation frameworks have been developed to support geocoding system design and performance assessment~\citep{goldberg2013evaluation,christen2002febrl}. Applied studies have primarily investigated the effects of geocoding quality on downstream spatial analyses~\citep{Zandbergen2012Error2SptialAnalysisSummaryofDifferentStuidies,goldberg2013evaluation}. Additionally, previous survey articles have covered geocoding from multiple angles, including sources of error and persistent challenges~\citep{goldberg2007text}, implications for health-related research~\citep{rushton2006geocodingforHealth}, and applications in Location-Based Services (LBS), particularly in the context of privacy protection~\citep{shin2012privacy}.

To inspire future advancements towards next-generation geocoding systems, this article begins by analyzing current demands for geocoding input and output as the objects to be achieved in Section \ref{Geocoding Input} and Section \ref{Geocoding Output}, respectively. We then break down a geocoding system into key components based on their roles in the workflow, examining their existing approaches, and proposing a set of research questions and hypothesis for investigation in Section \ref{Geocoding Components}. Finally, we identify research directions for the development of next-generation geocoding systems in Section \ref{sec:Opportunities in Geocoding}. Section \ref{Conclusions} concludes with a summary of this survey and future envisions. This object-driven and component-based perspective differentiates our review from the existing literature.

\section{Geocoding Input Demands}
\label{Geocoding Input}

The input for any geocoding system is a reflection of human perception and the subjective expression of a location. For example, the description for a POI might contains place abbreviations, in contrast to the public health domain, where inputs to geocoding are typically residential addresses. These input variations place significant demands on geocoding systems to handle them effectively. To create a clear picture of the inputs that geocoding systems must handle, we examine the sources of input heterogeneity and the challenges they introduce.

Table \ref{tab:geocoding_input_varies_by_scenarios} presents distinct location descriptions from the U.S. address system that have served as geocoding inputs across various studies and applications.
\begin{table}[ht]
\caption{Location descriptions for different application scenarios within the U.S. address system }
\label{tab:geocoding_input_varies_by_scenarios}
\centering
\begin{tabularx}{\textwidth}{@{}>{\RaggedRight\arraybackslash}X >{\RaggedRight\arraybackslash}X >{\RaggedRight\arraybackslash}X@{}} 
\toprule
\textbf{Application scenarios} & \textbf{Description patterns} & \textbf{Examples} \\ 
\midrule
Traffic accidents~\citep{levine1998carAccidentLocation} & Distance/Topology related to intersections or landmarks & 50 meters to the East of the intersection of Texas Ave. and Wellborn Rd. \\ 
\addlinespace
Disease Outbreaks~\citep{zinszer2010diseaseOutbreak} & Residential Postal Address & 209 Augsburg Ct., College Station TX 77843 \\ 
Crime Analysis~\citep{murray2011hybrid} & & \\ 
Exposure Risk Assessment~\citep{rushton2006geocodingforHealth} & & \\ 
\addlinespace
Location Based Services~\citep{tiwari2011survey} & Point of Interests (POI) Descriptions, Distance/Topology related to POIs & Eiffel Tower, Downtown Paris \\ 
\addlinespace
Delivery\footnotemark[1] & Post Office Box + 2-5 Digit Number & PO Box 123456 \\ 
\addlinespace
Named Entity Recognition (NER) from Articles~\citep{roberts2010toponym} & Administration Regions (e.g., city, state) & The Statue of Liberty is in \textit{New York Harbor in New York City}, in the United States. \\ 
\addlinespace
Indoor Navigation~\citep{di2005indoor} & Room Identifications + Building Identifications & Room SW160, Building E \\ 
\bottomrule
\end{tabularx}
\footnotetext[1]{\url{https://www.usps.com/manage/po-boxes.htm}}
\end{table}
These input variations can be explained by three factors. First, geocoding input describes where events or incidents often occur. Case reporting systems typically regulate the format of location descriptions during a data collection process. For instance, patients are required to provide their postal addresses upon hospital admission. Similarly, in Texas, the car crash report instructions \footnote{\url{https://ftp.dot.state.tx.us/pub/txdot-info/trf/crash_notifications/2023/cr100-v26.1.pdf}} mandate that each crash location be described in relation to the nearest intersections, referenced landmarks, or highway segments. Second, geocoding input reflects human cognition of locations~\citep{hu2019semanticPlaceName}. For example, the region termed ``downtown" does not physically exist on the earth but is defined by human activities, functions, and semantics~\citep{gao2017VagueRegions}. Such semantic descriptions frequently appear in daily communication, often containing abbreviations and aliases for locations~\citep{goldberg2013geocodingTechniques}. With the increasing prevalence of LBS such as ride-sharing and navigation in daily life~\citep{al2019towardsGeocodingSpatialExpressions, tiwari2011survey}, geocoding inputs now range from formatted postal addresses to vernacular points of interest (POI) descriptions~\citep{melo2017SurveyOfGeocodingTextualDocuments, goldberg2013geocodingTechniques}. Third, geocoding inputs stem from how humans artificially encode spatial areas. Different encoding schemes use combinations of alphabets and digits to represent spatial partitions, which ultimately serve as geocoding inputs. The most common encoding method is postal codes, assigned based on regional delivery needs~\citep{collins1998errorsZipcode, hurley2003post}. Similarly, room identification in buildings—encoding floor levels and segmentations based on arbitrary rules—functions as an input for indoor navigation scenarios.
\begin{table}
\centering
\caption{Examples of address formats in the Northern and Southern hemispheres countries}
\label{tab:address-examples}
\begin{tabular}{|l|p{10cm}|}
\hline
\textbf{Region / Country} & \textbf{Address components} \\
\hline
\multicolumn{2}{|c|}{Northern Hemisphere} \\
\hline
United States & [\textit{Street}] 1600 Pennsylvania Ave NW, [\textit{City}] Washington, [\textit{State}] DC [\textit{Postal Code}] 20500, [\textit{Country}] USA \\
\hline
France & [\textit{Street}] 10 Avenue des Champs-Élysées, [\textit{Postal Code}] 75008 [\textit{City}] Paris, [\textit{Country}] France \\
\hline
\multicolumn{2}{|c|}{Southern Hemisphere} \\
\hline
Brazil & [\textit{Street}] Av. Paulista, 1578 – [\textit{Neighborhood}] Bela Vista, [\textit{City}] São Paulo – [\textit{State}] SP, [\textit{Postal Code}] 01310-200, [\textit{Country}] Brazil \\
\hline
New Zealand & [\textit{Street}] 45 Cuba Street, [\textit{Suburb}] Te Aro, [\textit{City}] Wellington, [\textit{Region}] Wellington Region [\textit{Postal Code}] 6011, [\textit{Country}] New Zealand \\
\hline
\end{tabular}
\end{table}
In addition to differences in how people describe locations, geocoding inputs also vary by language and by the address system schema used in different countries. In Table~\ref{tab:address-examples}, we list four addresses from countries in the Northern and Southern hemisphere, respectively. The italicized text enclosed in brackets denotes the address components. For instance, the United States follows a sequence of \textit{Street} → \textit{City} → \textit{State} → \textit{Postal Code} → \textit{Country}, whereas France places the \textit{Postal Code} before the \textit{City}. In the Southern Hemisphere, Brazil inserts an intermediate \textit{Neighborhood} component between the \textit{Street} and \textit{City}, while New Zealand often includes a \textit{Suburb} and \textit{Region} before the \textit{Postal Code}. Thus, although the fundamental components of addresses are broadly similar, their order and the presence of these components largely depend on national conventions.

These variations in input format and content pose significant challenges for geocoding systems. Moreover, three key linguistic complexities further complicate the geocoding process. The first challenge is the semantics and ambiguity of words. Similar to semantic issues, ambiguity is a well-known problem in geocoding and Geographic Information Retrieval (GIR) domains~\citep{hu2018GIRChallenges, goldberg2013geocodingTechniques}. The ambiguity between street direction prefixes and street names frequently appears in geocoding practice. For example, "N Main St." could refer to a street named "North Main St." or "Main St." with a prefix of "North." The second challenge is that geocoding input is error-prone. A study has categorized input errors into two types: semantic errors and syntax errors~\citep{hutchinson2010abmGeocodingInpurError}. Semantic errors are the result of the substitution of words with similar meanings. For example, a user might input "Little River Rd" instead of "Small Creek Rd". Despite their similar meanings, geocoding systems struggle to distinguish them based solely on pronunciation or string similarities. Syntax errors often stem from habitual typing habits. For instance, the city name or the state name is often omitted in an address description~\citep{Christen2005hmmAddressParserInputErrors}. This omission can affect the reference feature matching process, as the same street name can be found in many cities. Besides missing address elements, typographic errors frequently occur in geocoding input and are found to be sensitive to keyboard layout~\citep{Kukich1992TypoCategroiesTypographicCognitive, Gundel1993ErrorCauseByKeyboard}. Studies have identified the most frequent typographic error patterns in geocoding input as follows~\citep{christen2002febrl, harries1999geocodingInputErrorsExample, Ranzijn2013AddressMatchingTechniques, Christen2005hmmAddressParserInputErrors, damerau1964singleErrorFreq}: (1) replacing one character, (2) deleting one character, (3) inserting one character, and (4) swapping two characters. The third challenge arises from differences in languages and address system schemas across regions, countries, and cultures~\citep{huang1996segmentationChineseLanguageCharteristizs}. The variation in the ordering of address components requires geocoding systems to go beyond a universal rule and adapt to country-specific conventions. Likewise, linguistic differences demand that geocoding systems correctly parse inputs written in different languages. For example, compared to U.S. postal addresses, Chinese addresses present the additional challenge of lacking white-space delimiters.
Recently, a benchmark dataset that simulates the aforementioned input typographical errors has been published and utilized to evaluate address parsing techniques ~\citep{yin2023chatgpt}. Nonetheless, further work is still needed to create additional benchmarks that encompass a wider spectrum of input variations.

In a nutshell, the heterogeneity of geocoding inputs underscores the need for next-generation geocoding systems to accommodate variations stemming from description practices, linguistic diversity, and differences in address system design. Common input challenges such as typographic errors and semantic ambiguity need to be appropriately addressed to improve output quality of geocoding systems.

\section{Geocoding Output Requirements}
\label{Geocoding Output}

In addition to returning spatial coordinates, geocoding systems typically produce supplementary attributes that describe the output ~\citep{goldberg2013evaluation}. Because geocoded data support critical spatial analysis workflows and applications, modern systems are expected to tailor these attributes to specific use cases \citep{jacquez2012research}. In this section, we summarize the roles of key output attributes, assess their limitations, and propose desirable enhancements. All these analysis collectively define the requirements for next-generation geocoding systems.

The key attributes of geocoding outputs fall into three categories: (1) geometric representation, (2) spatial accuracy, and (3) metadata description. These attributes shape the usability and reliability of geocoded data.

\subsection{Geometry representations} In terms of geometry, outputs from geocoding systems can take the form of points, polygons, or line segments, which represent buildings, parcels or areas, and street networks in reality ~\citep{zandbergen2008comparison}. These geometry representations are determined by input location description, the underlying matching algorithms, and reference datasets employed. For example, when city names are input into geocoding systems, the output could be either the polygon of the city boundary or the centroid of the city boundary. In cases where specific addresses (i.e., points) cannot be found in the reference datasets, commercial geocoding systems could revert to higher geographic levels (e.g., postal codes), generating the boundaries of postal codes as output ~\citep{goldberg2010candidateselectioncriteria}. This sensitivity to system-specific logic raises important questions and debates ~\citep{goldberg2007text}: Which geometry should be used as the output? Where is the optimal point to represent an entire city or postal code area? The answers to these questions often depend on the spatial resolution demands of subsequent studies. For instance, while using the centroid point of a county has been shown to be sufficient to represent disparities across a country ~\citep{zou2018TwitterDisaster}, it is not accurate enough when analyzing the accessibility of health service providers ~\citep{McDonald2017HealthAccessibilityGeocoding}. Therefore, it is crucial that any decision regarding geometric representation is clearly reflected in the geocoding output so that end users can make informed decisions about how to use the resulting geocodes. Yet, few existing systems provide this information.

\subsection{Spatial accuracy} Spatial accuracy in geocoding refers to the distance between geocoded locations and their corresponding ground-truth positions. It is the most critical metric for evaluating geocoding performance~\citep{goldberg2013evaluation,yin2019DLRooftopGeocoding,goldberg2010candidateselectioncriteria,zandbergen2008comparison,zhang2024mining}, significantly influencing subsequent spatial analyses~\citep{jacquez2012research,Zandbergen2012Error2SptialAnalysisSummaryofDifferentStuidies}. Previous studies indicate that spatial accuracy requirements and tolerances vary depending on the specific applications of geocoded data. For example, certain cancer risk assessments require spatial errors of less than 50 meters ~\citep{washburn199geocodedError50meters}, whereas exposure classification of Toxic Release Inventory (TRI) facilities can tolerate geocoding errors up to 1500 meters ~\citep{Zhan2006geocodingTWI1500meters}. For studies linking census information with individual data, only a small portion (i.e., 5\%) of geocoded data is affected by spatial error ~\citep{Kravets2007PointInPolygonError}, while street geocoding techniques can result in spatial errors exceeding 200 meters ~\citep{zandbergen2007exposure250MBiasResults}. Nevertheless, regardless of the analytical scale or approach, spatial accuracy significantly impacts study conclusions, particularly when geocoded locations fall near aggregation or classification boundaries~\citep{schootman2005geocodedPointInPolygon}. For example, when a geocode falls near the boundary of two house lots and its position is uncertain, providing information about the likelihood of association with each lot can be highly valuable. Thus, we argue that geocoding systems should provide a means to quantify and indicate the spatial relationship between output geocodes and nearby features. We will elaborate on this requirement in Section \ref{sec:Metadata}.

\subsection{Metadata description}\label{sec:Metadata} Descriptive metadata for geocoding systems are metrics that describe the quality of each individual or the entire geocoded output collection. The capability of reporting metadata is considered as a crucial criteria to differentiate geocoding systems ~\citep{goldberg2013evaluation}. Common geocode output metadata descriptions from different geocoding systems are listed in Table \ref{tab:metadata_description_for_geocodes}. These descriptions can be grouped as spatial and non-spatial.
\begin{table}[ht]
\caption{Metadata Description for Geocodes}
\label{tab:metadata_description_for_geocodes}
\centering
\begin{tabularx}{\textwidth}{@{}l X c@{}}
\toprule
\textbf{Name} & \textbf{Function} & \textbf{Is spatial description} \\ 
\midrule
Match rate & Percentage that the system is able to find a matched geocode & N \\ 
Match score & Customized score to indicate match quality for an output geocode & N \\ 
Match level & Geographic level that the system is able to match for an output geocode & Y \\ 
Geometry & Geometry of the output geocode & Y \\ 
\bottomrule
\end{tabularx}
\end{table}
Generally, spatial descriptions are preferred over non-spatial ones, non-spatial descriptions lack the information needed to accurately evaluate the quality of geocodes. For example, given a set of geocoding inputs, depending on the underlying algorithm, a system may report high match rates (a non-spatial metric) by reverting a less spatial accuracy level to find a match. Consequently, relying solely on the match rate without considering the spatial accuracy can negatively impact the potential usage of output geocodes, particularly when one needs to select an appropriate system for data processing. 

To date, research has primarily focused on evaluating collections of geocoded outputs rather than individual results. This includes modeling spatial error across different environments \citep{Cayo2003ResidualUrbanRural}, geocoding techniques \citep{roongpiboonsopit2010quality}, and reference datasets \citep{vieira2008accuracy,zandbergen2011influence}. Yet these models offer limited insight into the quality of individual geocodes. Recent research has proposed using spatial uncertainties—the area within which the true location is likely located—as \textit{new spatial metadata} to describe geocode quality~\citep{goldberg2010towardQuantitativeGeocodeAccuracy}. As aforementioned, when output geocodes fall close to the boundaries of areas such as parcels or counties, quantifying the likelihood that the geocode belongs to each area is of importance. However, few systems currently include this metadata, as it requires integrating a quantification algorithm into existing geocoding workflow to calculate these uncertainties between reference data and candidate outputs~\citep{goldberg2010candidateselectioncriteria,goldberg2010towardQuantitativeGeocodeAccuracy}.

Given the facts presented above, we can conclude that the choice of geometric representation and spatial accuracy level should be closely tied to the specific needs of subsequent application scenarios. In addition, the spatial metadata associated with individual geocodes—or with a set of geocodes—can be critical for downstream tasks, helping users determine whether the outputs are suitable and what potential biases may arise from their usage. Accordingly, we propose the following  new requirements that next-generation geocoding systems should meet: 
\begin{itemize}
\item providing a means to illustrate the rationale behind choosing one geometric representation for output over others, especially when spatial accuracy is sacrificed when attempting to identify a geocoded. 
\item reporting the necessary spatial uncertainty for each individual geocode in relation to nearby features, particularly when the geocode lies close to boundaries or other spatially adjacent features. 
\end{itemize}

\section{The State of Geocoding Components}
\label{Geocoding Components}
Having defined the standards for geocoding inputs and outputs, we now turn to the geocoding process, which accepts these inputs and generates the corresponding outputs. Geocoding is a complex workflow involving natural language processing, probabilistic retrieval, spatial feature matching, and candidate selection \citep{zandbergen2008comparison}. To identify solutions that meet the high standards defined above, it is essential to examine each component of this workflow in greater depth and to pinpoint the gaps that may still remain. Thus, in this section, we first decompose the geocoding process into distinct and manageable components. We then describe the functionality and current state of each component in detail. Finally, we outline the challenges and the future research directions for each component when building the next-generation geocoding systems.

\subsection{Geocoding Workflow Decomposition}

\begin{figure}[ht]
  \includegraphics[width=\textwidth]{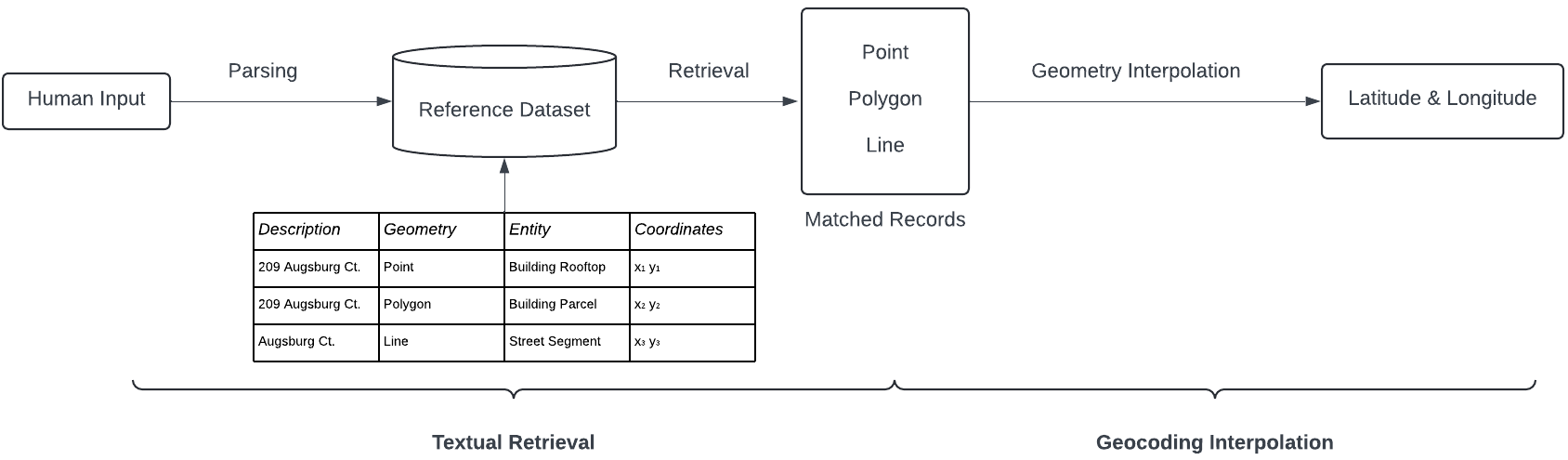}
  \caption{Geocoding workflow}
  \label{fig:workflow}
\end{figure}

Figure \ref{fig:workflow} illustrates the geocoding workflow, which can be broadly divided into two main parts: \textit{geocoding textual retrieval} and \textit{geocoding interpolation}. The goal of the geocoding textual retrieval is to identify the record in the reference dataset that most closely corresponds to the input location description, whereas the geocoding interpolation derives the final output geocodes from the retrieved records based on their spatial relationships. During the textual retrieval phase, the geocoding system parses the input description into system-interpretable components, constructs a query string, searches the reference datasets, and ranks the retrieved candidates according to their matching scores. Therefore, we further divide this phase into three closely linked components: (1) data storage, (2) input parsing, and (3) retrieval, as the retrieval algorithm dictates how reference data should be segmented and stored, and how input address description should be parsed. The second part: geocoding interpolation can be considered a standalone component that focuses solely on deriving the final output coordinates using the geometry and spatial relationship of the retrieved reference data. Based on this decomposition, we now elaborate on the current state of each of these four components in the following sections.
 
\subsection{Geocoding Textual Retrieval}
\label{Geocoding Textual Retrieval}

As discussed earlier, data storage, input parsing, and retrieval are tightly coupled within the textual retrieval phase. Table \ref{tab:component_variations} summarizes the variations in these three components, grouped by their implementation methodologies, reported in existing research. We examine each of these variations in greater detail below.

\begin{table}[ht]
\caption{Summary of Variations in Geocoding Textual Retrieval Components}
\label{tab:component_variations}
\centering
\renewcommand{\arraystretch}{1.3} 
\setlength{\tabcolsep}{10pt} 
\begin{tabularx}{\textwidth}{|l|X|}
\hline
\textbf{Component}       & \textbf{Variations} \\ \hline
\textbf{Data Storage} & 
\begin{itemize}[leftmargin=*, labelsep=10pt]
    \item \textbf{Segmentation type:} Address component / Non-address component 
    \item \textbf{Data representation type:} Phonetic-based / (Sub) String-based / Dense vector-based 
    \item \textbf{Storage media type:} SQL-based / NoSQL-based / Binary file-based
\end{itemize}
\\ \hline
\textbf{Input parsing}           & 
\begin{itemize}[leftmargin=*, labelsep=10pt]
    \item Rule-based 
    \item Statistic-based
    \item Neural network-based 
\end{itemize}
\\ \hline
\textbf{Retrieval}          & 
\begin{itemize}[leftmargin=*, labelsep=10pt]
    \item Rule-based 
    \item Probabilistic-based
    \item Machine learning-based
\end{itemize}
\\ \hline
\end{tabularx}
\end{table}

\begin{figure}[ht]
  \includegraphics[width=\textwidth]{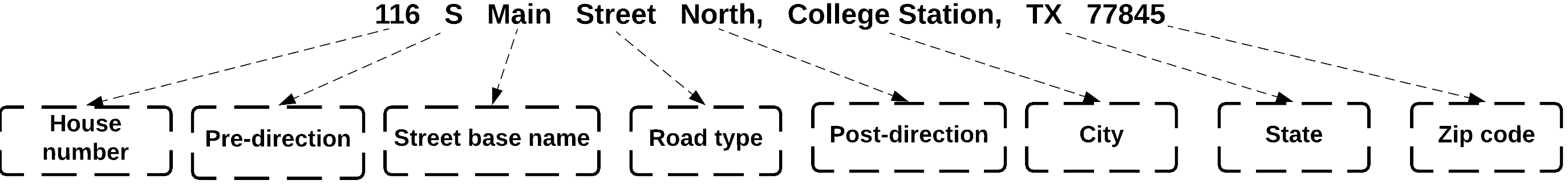}
  \caption{The standard address components for U.S. addresses}
  \label{fig:USPS address components}
\end{figure}

\subsubsection{Data storage}
\label{sec:Data Storage}

Data storage, functioning much like a digital gazetteer that the system consults, determines how the reference dataset is organized, accessed, and retrieved. It can be characterized from three perspectives: segmentation type, data representation type, and storage media.

\textbf{Segmentation type} refers to how we divide an entire address description record into several blocks for further processing and storage in the database. Here, we only provide a high-level abstraction of how segmentation of an address record is determined, as some string similarity algorithms—discussed later in this section—may further refine certain portions of the address segmentation. At a high level, segmentation types can be categorized into address component-based and non-address component-based types. Address component-based segmentation divides each address record based on standard address components defined by a country's address system. Figure \ref{fig:USPS address components} presents an example of the standard address components defined in the United States. In contrast, non–address-component-based segmentation permits arbitrary combinations of standard components. For instance, forming a street-level description from a pre-direction, base street name, road type, and post-direction. The choice of segmentation type is controlled by the retrieval methods employed in a geocoding system, which will be discussed in detail in Section \ref{sec:retreval}. For example, if a geocoding system utilizes the \textit{Per-attribute Score Ranking} method \citep{goldberg2010candidateselectioncriteria}, which gives each address component a weighting score, each address record is more likely to be segmented according to individual address components. If combining house number, street base name, and street type can improve the quality of retrieved matching candidates, these address components then should be merged, necessitating the use of non-address component segmentation. In addition to standard address components, studies have introduced new fields to improve retrieval quality. For example, alias fields for city and state have been incorporated to account for abbreviations or outdated names commonly found in short messages ~\citep{zhang2014geocodingTwitter}.

\textbf{Data representation type} refers to how each address record in the reference datasets is represented. It can be classified into three groups: (1) phonetic, (2) string, and (3) dense vector, each associated with a specific set of matching algorithms designed to enhance the accuracy of record matching within the dataset.

In the phonetic-based group, address descriptions are converted into alphanumeric-encoded strings using phonetic algorithms to handle misspellings based on pronunciation \citep{toutanova2002pronunciation}. A commonly used algorithm in geocoding systems is Soundex \citep{zandbergen2008comparison, goldberg2010candidateselectioncriteria}. However, it is limited to English and sensitive to the position of typos \citep{zandbergen2008comparison}. To address these issues, the Double-Metaphone algorithm was introduced to accommodate non-English languages \citep{philips2000doublemetaphone, shresthaa2011DMGeocoding}. However, it still faces challenges, such as different words sharing the same encoding \citep{lisbach2013linguistic}.

The string-based group represents each address record as a string and employs string-based similarity algorithms to identify matching candidates. Various string-based similarity algorithms have been used in geocoding systems \citep{Ranzijn2013AddressMatchingTechniques, yin2021systematic, santos2018learning}. Depending on the particular algorithm, strings in each field may undergo modifications. For example, before calculating similarity, token-based algorithms such as \textit{q}-gram algorithms \citep{li200qGramFilterByThreshold} first break strings into smaller segments, a pad version of \textit{q}-gram algorithm \citep{keskustalo2003PadVersionQGramIsBetter} add special padding characters to original strings. These string-based algorithms are typically used to match street-level descriptions and city names. Recently, \citet{yin2021systematic} applied \textit{q}-gram methods to postal codes, outperforming the traditional approach of matching only the first three digits. Thus, the ultimate representation of each address record is closely linked to the choice of string-based similarity algorithms for each address field.  

The dense vector-based representation is an emerging string encoding approach, driven by recent advances in NLP and vector databases \citep{mikolov2013efficient, vaswani2017attention, han2023comprehensive}. This method converts address descriptions into dense vectors, effectively capturing semantic relationships between words to mitigate potential semantic errors in input data. Existing geocoding literature has explored their retrieval performance of various word embedding techniques, ranging from Word2Vec to transformers \citep{lin2020W2VforGeocoding, duarte2023transformergeocoding, comber2019geocodingMatchWord2VecCDF}.

It is important to note that not all address segments need to be represented in the same format, as the representation type is closely linked to the matching algorithms that tend to be used for this segment. 
The choice of algorithms to be used to match each field should be evaluated on a case-by-case basis. This is because the performance of these matching algorithms depends heavily on the content and types of human errors present in each field. Additionally, the overall matching quality is influenced by matching/non-matching classifiers and the way a classifier utilizes the calculated similarity values, which will be discussed further in Section \ref{sec:retreval}.

\textbf{Storage media type} refers to where and how reference datasets are stored, which can significantly impact the implementation of a geocoding system. Different databases offer varying levels of support for data models and query languages, influencing system performance and functionality. However, limited literature have conducted experiments on the comparison of geocoding performance resulting from a database or particular implementation. One reason could be proprietary concerns, especially for commercial geocoding systems. In this section, we provide a brief overview of storage media type used by geocoding systems, encouraging further research to explore this area in more depth. Historically, SQL-based databases have been dominant in geocoding fields ~\citep{Goldberg2011ImprovingMatchRateVaryingBlock, Xu2012GeocodingBillionAddress}. However, the key advantage of SQL-based databases—their ability to join fields from different tables based on specific conditions—is mainly used to process reference datasets rather than to query matching candidates ~\citep{goldberg2008geocodingbest}. To this end, recent researches have examined the scalability for NoSQL-based database, with the document-oriented and easy data partitioning design ~\citep{clemens2020geocodingPhdDissertation, knoblock2017ElasticSpatialExampleGeocoding, guo2020geocodingDatabase}. Since geocoding involves searching within a reference dataset, which is treated as a fixed resource pool, some geocoding systems convert the entire reference dataset into a binary file for more efficient search processing \footnote{\url{https://help.precisely.com/r/Spectrum/global_geocoding/23.1/en-US/Spectrum-Global-Geocoding-Guide/Custom-Dataset-Builder/Building-a-Custom-Dataset-for-Geocoding-and-Typeahead}}. Due to potential commercial interests, limited research has been published on the performance of geocoding systems that utilize binary files as their reference dataset format. Thus, establishing benchmarks for geocoding performance across different storage media could provide valuable insight about which storage media are the most efficient.

\subsubsection{Input parsing}

Geocoding address parsing (referred to as geocoding parsing) is a key focus in geocoding research, as it governs the construction of the query string in the downstream geocoding workflow and therefore impacts the quality of the resulting geocodes. This process involves segmenting the input location description and assigning a label to each resulting token~\citep{boscoe2008science}. The specific tokens used should depend on the geocoding system's retrieval algorithm, namely how the reference data is segmented and how match candidates are ranked. Standard address components, as shown in Figure \ref{fig:USPS address components}, are commonly used as target labels assigned to input address descriptions. It's important to distinguish geocoding parsing from more general address parsers, which focus on identifying and extracting location descriptions from text and assigning entity labels (e.g., \textit{LOC}). This broader task, known as toponym recognition or text geocoding, doesn't require parsing the finer elements of addresses. Recent surveys by \citet{hu2023location} and \citet{melo2017SurveyOfGeocodingTextualDocuments} provide comprehensive reviews on toponym recognition. However, to the best of the authors' knowledge, few surveys have been published that thoroughly and specifically review geocoding address parsing techniques.

Existing geocoding parsing solutions can be categorized into three approaches: (1) rule-based, (2) statistical-based, and (3) neural network-based. Compared to statistical-based and neural network-based approaches, rule-based approaches take a ``manual" approach, as they utilize the structure and hierarchy of defined address system schemas to create logic to assign labels to input address descriptions ~\citep{goldberg2013geocodingTechniques}. For example, in the U.S. address system, transitions between address components- house numbers followed by street level descriptions and 5-digit postal code typicall appears at the end of address description, are predictable. Current research on rule-based address parsers focuses on (1) abstracting address system hierarchies and (2) applying search algorithms (e.g., sliding window and heuristic search) and string-matching methods (e.g., edit distance) along with gazetteers to process addresses from various countries, including China, U.S., Turkey, and Japan ~\citep{zhnag2010ruleBaseChineseAddress,goldberg2008geocodingbest,liu2002JapanLexicon,matci2018address}. However, limitations of rule-based methods have also been pinpointed. Address systems differ across countries, cultures, and languages, requiring a unique abstraction of each country's hierarchy. In countries without a well-defined address system, rule-based methods are not applicable. Additionally, these methods rely on lexicons to recognize address components, making them prone to ``Out of Vocabulary" issues when user inputs vary in quality or description ~\citep{li2018hybridChineseAddressSegmentions}.

To address these challenges, both statistical and neural network-based approaches have been developed, utilizing annotated training data to ``learn" transitions between possible address components. Among the statistical methods, two popular models—Hidden Markov Models (HMMs) and Conditional Random Fields (CRFs)—are frequently applied for address parsing tasks~\citep{Christen2005hmmAddressParserInputErrors,sun2017chineseCDFNER}. HMMs aim to maximize the joint probability between the observed sequence and states, assuming independence between each word and state, while CRFs improve on this by maximizing the conditional probability of labels given an observation sequence, using undirected graphical models. Initially used for NER tasks, both models have been extended to annotate address components. While most studies evaluate their performance across various countries' address formats, further research seeks to enhance performance by integrating address system rules~\citep{li2018hybridChineseAddressSegmentions}, stochastic regular grammar-based scoring functions~\citep{wang2016CRFAddressParser}, and by reducing the amount of training data required~\citep{craig2019BingmapParserConcatenateW2Vfeatures}.

Since address descriptions can be viewed as a sequence of data (i.e., text), it falls into the scope of Recurrent Neural Networks (RNNs), which aim to consider both the input for the current elements and preceding elements for each element in sequences. Among these RNN model architectures, Long Short-term Memory Networks (LSTMs), which introduce a memory-cell to capture long-range dependencies ~\citep{hochreiter1997LSTM}, have shown advantages over traditional RNN models. Variations of LSTM models, including Bidirectional LSTM, attention-based LSTM, and hybrid LSTM-CRF models, have been developed and evaluated in NER tasks \citep{lample2016LSTMCRf,zhou2016attentionLSTM,chiu2016bidirectionalLSTM}. These LSTM-based architectures have inspired significant research into place name (LOC tag) recognition in various contexts \citep{wang2020neurotpr, cardoso2019using}. As noted earlier in this section, while address labeling and place name recognition both fall within the NER field and share certain similarities, their contextual sequences differ. Consequently, the performance of these models may not directly transfer to address labeling recognition and might require additional fine-tuning or evaluation. Although few studies have explicitly evaluated sequential neural networks for address labeling, these models have shown strong results. For example, \cite{yassine2021universalAddressParsing} used subword embeddings and an LSTM-based embedding layer with a sequence-to-sequence (Seq2Seq) model to build an address parser, achieving state-of-the-art performance across addresses from 20 countries.

More recently, LLMs, such as BERT \citep{devlin2018bert} and GPT (Generative Pre-trained Transformer) \citep{brown2020language}, have revolutionized the field of NLP. These LLMs are pre-trained on vast amounts of unstructured text data, enabling them to learn rich linguistic representations and capture complex language patterns. One way to fine-tune LLMs to the specific address parsing task. For example, \cite{li2023enex} proposed a BERT-based model for address parsing called AddressBERT. They fine-tuned BERT on a large dataset of labeled address strings and achieved state-of-the-art performance in parsing addresses from multiple countries and languages. The model learns to identify and extract relevant address components based on the learned patterns and context. The other way is to leverage prompt engineering with generative AI. For instance, \cite{yin2023chatgpt} benchmarked the performance of ChatGPT, with a specific prompt with just three address labels, using 230,000 U.S. address descriptions containing various input errors. The study showed that ChatGPT produced promising parsing results, compared to LSTM-based address parsers. Similarly, \cite{huang2024geoagent} employed a prompt-based approach using ChatLM to parse Chinese addresses.  

\subsubsection{Retrieval}
\label{sec:retreval}

Retrieval aims to retrieve the most relevant matching candidates from the reference dataset based on the input address description. This process can be generalized by the following equation.
\begin{equation}
\textit{Matching score} = \textit{Classifier(Similarity)}
\end{equation}
where the retrieval process first identifies a set of matching candidates using string similarity methods, and it then applies a classifier function, which takes the calculated similarity value for each candidate as the input, to output matching scores that rank the candidates. Table \ref{tab:comparison_retrival_techniques} lists some distinct combinations of similarity calculation methods and classifiers used in existing geocoding literature.
\begin{table}[h!]
\centering
\caption{Comparison of Retrieval Techniques in Geocoding Systems}
\label{tab:comparison_retrival_techniques}
\begin{tabularx}{\textwidth}{|X|X|X|X|}
\hline
\textbf{Geocoding System} & \textbf{Similarity} & \textbf{Classifier} & \textbf{Category} \\ \hline
\cite{goldberg2010candidateselectioncriteria} & Phonetic and string similarity & Predefined weight & Deterministic \\ \hline
\cite{li2014region} & Phonetic and string similarity & Probabilistic weight & Probabilistic \\ \hline
    \cite{comber2019geocodingMatchWord2VecCDF} & Vector similarity & Random Forest + XGBoost & Machine learning \\ \hline
\cite{rezaei2024address} & String similarity & SVM & Machine learning \\ \hline
\cite{lin2020W2VforGeocoding} & Vector similarity & ESIM & Machine learning \\ \hline
\cite{duarte2023improving} & Pre-train model embedding & K-NN Similarity & Machine learning \\ \hline
\end{tabularx}
\end{table}
It's worthy noting that all these similarity methods and classifier could be combined randomly. The details of these similarity methods have been discussed in Section \ref{sec:Data Storage}, as the choice of similarity calculation methods also has impacts on how data are segmented and stored in a database. Thus, in this section, we focus on classifiers utilized by different geocoding systems, since the existing literature often used the type of classifier to differentiate the types of geocoding systems: deterministic and probabilistic \citep{boscoe2008science, goldberg2013geocodingTechniques}. We define a new category of \textit{machine learning} for geocoding systems that employ a machine learning based classifier. 

A deterministic classifier typically uses a predefined scoring mechanism to determine if matched candidates are found and then rank these candidates. For example, \cite{goldberg2010candidateselectioncriteria} proposed a \textit{Per-attribute Score Ranking} method to evaluate geocode candidates by assigning scores based on the match status of individual address components, each pre-assigned a weight reflecting its importance in determining a match. In addition to these address component scores, the method requires a minimum matching score to distinguish between match and non-match statuses. These empirically-determined score mechanisms limit its applicability to (1) address descriptions from other address systems and (2) matching fields that differ from the scoring field.
 
Unlike a deterministic classifier, which uses empirically determined weights for each matching field, a probabilistic classifier assigns weights proportional to the frequency of each field in the reference datasets. A classic weight calculation method is known as the Fellegi and Sunter model \citep{fellegi1969theoryFSmodel}, which is widely used in the domain of record linkage as follows \citep{sayers2015probabilisticRecordLinkageReview, gu2003recordLinkageReview}.
\begin{equation}
\label{eqn:probWeightGeocoding}
W_{i} = \ln{\frac{m}{\mu}} = \ln{\frac{P(\gamma_{i} \mid M)}{P(\gamma_{i}\mid U)}}  \propto \ln{\frac{N}{n}}
\end{equation}
where $W_{i}$ denotes the weight for the $i$th field, $\gamma_{i}$ denotes the agreement of the compared field $\gamma_{i}$, 
$m$ denotes the probability that the compared field agrees given that it is a true match $M$ and $\mu$ denotes the probability that the compared field agrees given that it is not a true match $U$. In geocoding practice ~\citep{sayers2015probabilisticRecordLinkageReview, boscoe2008science}, $P(\gamma_{i} \mid M)$ is assumed to be very close to 1, and $P(\gamma_{i} \mid U)$ is typically approximated by the fraction of the number of records with the field $\gamma_{i}$ and the number of records in collections. Later, researchers started to adopt full-text document search engines (e.g., Elasticsearch \footnote{\url{https://www.elastic.co/elasticsearch}}) to build geocoding systems ~\citep{clemens2015geocodingESwithAddrTokenDropped, knoblock2017ElasticSpatialExampleGeocoding}. Thus, two weight calculation methods: \textit{Term Frequency - Inverse Document Frequency} (Equation \ref{eqn:tfidf}) and Best Match 25 (Equation \ref{eqn:bm25}), which are commonly used in IR, have been used as the classifier to rank geocoding match candidates.
\begin{equation}
\label{eqn:tfidf}
TF\text{-}IDF(t, d, D) = TF(t, d) \cdot IDF(t, D)
\end{equation}

\begin{equation}
\label{eqn:bm25}
BM25(t, d, D) = \sum_{t \in q} IDF(t) \cdot \frac{f_t \cdot (k_1 + 1)}{f_t + k_1 \cdot \left(1 - b + b \cdot \frac{|d|}{\text{avgdl}}\right)}
\end{equation}

While TF-IDF and BM25 differ in their weight calculation methods, both prioritize terms that appear less frequently in reference datasets. In TF-IDF, the inverse document frequency (IDF) is calculated as \( IDF = \log \frac{N_D}{1 + n_t} \), where \( N_D \) is the total number of documents in the corpus \( D \), and \( n_t \) is the number of documents that contain the term \( t \). In contrast, BM25 incorporates the absence of the term in the corpus and applies smoothing as follows: \( IDF = \log \frac{N_D - n_t + 0.5}{n_t + 0.5} \). Regarding term frequency, TF-IDF directly uses the raw term frequency without adjustment, while BM25 introduces saturation to limit the dominance of frequently occurring terms, thereby preventing them from disproportionately influencing the relevance score. 

A machine learning classifier learns thresholds to discriminate similarity values to each comparison on the field in an address description using annotated matched and unmatched address pairs using machine learning algorithms. As shown in Table \ref{tab:comparison_retrival_techniques}, these algorithms range from classical machine learning classification models to neural network-based approaches. 
\cite{comber2019geocodingMatchWord2VecCDF} combined Random Forest and XGBoost classifiers, using Word2Vec embeddings as input to match U.K. addresses. This ensemble approach outperformed each classifier individually. \cite{rezaei2024address} compared the performance of linear and nonlinear Support Vector Machines (SVMs) for address matching in Iran, demonstrating that nonlinear SVMs outperformed their linear counterparts. This result is in line with the findings of \cite{comber2019geocodingMatchWord2VecCDF}, which also reported superior performance of nonlinear classifiers compared to linear ones. With recent advances in NLP, \cite{lin2020W2VforGeocoding} trained, using a corpus of Chinese addresses, a Word2Vec embedding model, and an Enhanced Sequential Inference Model (ESIM) to predict match and unmatch labels for two addresses, outperforming traditional classifiers like Random Forest and SVM with the same Word2Vec embeddings.
Instead of obtaining word embedding for segmented address components, \cite{duarte2023improving} fine-tuned a pre-trained DistilBERT model to generate vector representations for a whole address description. They then retrieved the 10 most similar addresses and re-ranked them to derive the final output by processing these pairs through a transformer encoder. This approach achieved higher matching accuracy and shorter inference times compared to the BM25 method to match Portuguese addresses.

To date, all three types of classifiers have been utilized in geocoding systems. To make an informed choice, different classifiers should be systematically evaluated on a unified data set that closely reflects the quality and characteristics of the inputs the system is expected to encounter in production.

\subsection{Geocoding Interpolation}

The purpose of geocoding interpolation is to derive the final output coordinates based on the geometric representations of the reference data given an input. Existing interpolation methods, which are tied to the geometry of reference datasets, can be categorized into three types: (1) linear interpolation, (2) area interpolation, and (3) point interpolation.

\textbf{Linear interpolation}. Linear interpolation calculates the output coordinates by linearly scaling the distance between known reference points (e.g., start/end point of a street segmentation). This method assumes that the change between the points follows a straight line, allowing the output coordinates to be estimated proportionally based on the input's relative position along the line. While this is the most commonly used geocoding interpolation method due to the extensive coverage provided by line-based reference models (e.g., street centerlines), it may yield the least spatially accurate geocodes ~\citep{boscoe2008science}. The lower spatial accuracy in linear interpolation stems from several key assumptions, collectively known as the premise of parcel homogeneity ~\citep{dearwent2001locationalUncertaintyInGeocoding}. These assumptions include: (1) the existence of addresses, where parcel counts are solely based on house number ranges; (2) the uniform lot assumption, which presumes that all parcels on a street are of equal size; and (3) the parcel extension assumption, which assumes that parcels completely occupy a street without gaps at corners ~\citep{goldberg2008geocodingbest}. In common linear reference datasets, such as USPS TIGER\footnote{\url{https://www2.census.gov/geo/pdfs/maps-data/data/tiger/tgrshp2017/TGRSHP2017_TechDoc_Ch4.pdf}}, house numbers are often arbitrarily assigned to address ranges for privacy protection, even if they do not exist in reality. Consequently, the linear interpolation algorithm operates on the flawed concept of "coordinate difference per house number difference" without additional contextual information. Furthermore, geocoding systems often predefine the offset distance from a street centerline to a building without adjusting for local variations, introducing additional errors into the final output ~\citep{grubesic2004assessing}.

\textbf{Area interpolation}. Area interpolation uses polygon-based reference datasets to determine a geocoding output point from a given polygon. The main challenge with this method is accurately placing the point to best represent the entire area. Typically, there are three approaches to generating the output point: (1) bounding box, (2) centroid, and (3) weighted centroid ~\citep{goldberg2013geocodingTechniques}. However, all these methods assume that the derived point accurately represents the building's location within the polygon. While this assumption holds for small parcels, it often breaks down for larger parcels, especially in rural areas ~\citep{Cayo2003}. Additionally, for irregularly shaped parcels, such as those in "L" or "S" shapes, the centroid may not lie within the parcel boundaries ~\citep{GOLDBERG201239}. 

\textbf{Point interpolation}. Address points, by definition, represent the actual locations of buildings, eliminating the need for interpolation to derive the final geocode ~\citep{zandbergen2008comparison}. As a result, this geocoding method is considered to produce the most accurate geocodes. However, reference datasets for address points, which are typically generated either by deriving centroids from building footprints or parcels, or through field collection using Global Positioning System (GPS) technology, are resource-intensive in terms of time and effort ~\citep{zandbergen2008comparison,GOLDBERG201239}. Thus, such datasets often have limited geographic coverage.

In summary, the spatial accuracy of these three methods is mainly constrained by the available knowledge of building locations and distributions during geocoding interpolations. To address this limitation, several studies attempt to incorporate auxiliary data source, such as online records of building counts and parcel sizes~\citep{Bakshi:2004:EOS:1032222.1032251}, population density ~\citep{Schultz_2007}, and building footprints detected from satellite imagery ~\citep{yin2019DLRooftopGeocoding}. Nonetheless, each of these interpolation approaches exhibits significant limitations when auxiliary data sources are limited or absent.

\subsection{Future Directions for Geocoding Components}
Based on the analysis of the current state of each geocoding component, we summarize the key challenges and propose paired experimental questions and hypotheses for each component to provoke further investigation.

For \textbf{Data Storage} component, the main challenge is how to make the correct choices to deliver the most effective storage solution that yields high retrieval quality and throughput. To find the answers, we propose the following research questions and hypothesis for investigation.
\begin{itemize}
    \vspace{0.25em}
    \item \textit{Q1:} Which address record segmentation strategy (i.e., address component-based type vs. address component type) should be used for storage? \\
    \textit{H1:} Segmenting the address description according to address elements gives more flexibility and leads to better retrieval results.
    \vspace{0.5em}
    \item \textit{Q2:} Which data representation formats (i.e., phonetic encoding, string, dense vector) yield the highest retrieval performance across diverse inputs?\\
    \textit{H2:} Dense vector representations outperform string-based and phonetic formats for string-based address elements in general, whereas string-based formats still work well in numeric only address elements such as house number or postal code.
    especially for noisy or ambiguous inputs.
    \vspace{0.5em}
    \item \textit{Q3:} Which storage media should be used to maximize the system's retrieval throughput?\\
    \textit{H3:} High-performance storage (e.g., in-memory binary file) could potentially increase throughput compared to traditional database techniques.
    \vspace{0.25em}
\end{itemize}
The first question focuses on the address record segmentation strategy that should be used for storage. Based on the analysis in Section \ref{sec:Data Storage}, the segmentation type is highly tied to the parsing results and classification approaches. Thus, we argue that segmenting the address description according to standardized address elements can maximize flexibility since the query build algorithm can combine any of the standardized address elements as needed. The second question addresses the choice of representation formats appropriate for each address segmentation type. Given the fact that geocoding input could have both syntactic and semantic errors, we argue that the choice of representation format should be tailored to each address segmentation. In general, dense vector representations work better on the string-based on address elements such as street- or city-level descriptions, whereas the string representation is more efficient for some numeric only fields by the usage of fuzzy string matching techniques. It is noted that a large portion of this question can be answered by the classifier. The last question concerns how does storage media impact the system's performance. Given the geocoding reference dataset is considered a fixed and structured data source, if one can compress the dataset and hold them in memory, the query throughput can be significantly increased.
All three research questions can be evaluated using the same set of metrics. Specifically, recall and precision can be used to assess retrieval quality, while the number of geocodes processed per second can be used to measure system throughput.

For the \textbf{Input Parsing} component, the main challenge lies in training an address parser that can fully handle the complexity of geocoding inputs. Given the success of modern NLP methods, neural network–based approaches are generally more capable than rule-based or statistical approaches in addressing this complexity. To this end, we propose three research questions to investigate this issue.
\begin{itemize}
    \vspace{0.25em}
    \item \textit{Q1:} Which neural network architecture should be utilized to train an efficient address parser in terms of throughput and parsing quality? \\
    \textit{H1:} Transformer-based address parsers offer greater capability than RNN-based neural models but incur higher computational costs during both training and inference.
    \vspace{0.5em}
    \item \textit{Q2:} For the same neural network architecture, does adding non-address short-text context alongside the location description enhance training outcome? \\
    \textit{H2:} Embedding the location description with non-address short-text context can lead to improved training performance.
    \vspace{0.5em}
    \item \textit{Q3:} Can a prompt-based address parser backed by generative models outperform the transformer-based address parser?\\
    \textit{H3:} Prompt-based address parser requires more efforts to fine-tune a prompt to guide generative models to extract all possible address elements.
    \vspace{0.25em}
\end{itemize}
The first two questions concern how to train a more robust address parsers. Regarding training data, most existing work relies solely on datasets containing address descriptions, without incorporating additional contextual information. It would be meaningful to examine whether combining fine-grained address description annotations with non-address context could improve training outcomes. \cite{sun2025galloc} developed a platform to annotate location descriptions in disaster-related text messages, which may provide useful initial labeled data. However, training an address parser ultimately requires fine-grained annotations for each address element, as defined in Figure \ref{fig:USPS address components}. Thus, obtaining such fine-grained training data becomes an essential prerequisite. As for the model architecture, although location descriptions are relatively short, RNN-based models may still experience information loss due to long-sequence dependencies—an issue that can be amplified when the training corpus includes text beyond standard address descriptions. In contrast, transformer-based architectures encode the entire sequence simultaneously and employ multi-head attention to capture intra-sequence relationships more effectively. However, these architectures are also more computationally intensive, leading to more financial costs in the training time and the preparation of the annotated data. These factors underscore the need to evaluate which model is appropriate balanced between model performance and efficiency. Furthermore, incorporating location descriptions from multiple languages would offer valuable insights into how well each model architecture generalizes across linguistic contexts. The last question attempts to evaluate whether or not prompt engineering backed by LLMs can outperform a neural network address parser. We argue that prompt engineering might require more effort to crafted the prompt templates. For example, \cite{hu2023geo} discovered that providing a GPT model with knowledge about location descriptions improves its ability to extract LOC tags and certain standard address elements. Given the fact that more and more foundational LLMs are released, it would be interesting to conduct evaluation comparisons to see if a prompt-based address parser can outperform a neural-network based address parser in terms of parsing quality and the processing latency.

For the \textbf{retrieval} component, a major challenge is to determine which classifier algorithms and representation formats offer the greatest robustness, particularly given the variety of retrieval models currently deployed in production systems, as discussed earlier. Thus, we propose the following research question.  
\begin{itemize}
    \vspace{0.25em}
    \item \textit{Q:} Which combinations of classifier algorithms and representation formats provide the most robust retrieval performance across heterogeneous and noisy input scenarios? \\
      \textit{H:} Machine learning-based retrieval methods that incorporate learned or hybrid representations (e.g., dense vectors combined with phonetic or string-based features) will achieve higher robustness and retrieval accuracy across diverse input conditions than approaches relying solely on a single representation. 
    \vspace{0.25em}
\end{itemize}
The answer to this question is critical, as it directly impacts the choice of representation formats used to store address records in the reference dataset. We argue that machine learning–based retrieval models generally offer greater robustness because they can adapt to specific geocoding scenarios more effectively than rule-based or probabilistic methods. Additionally, hybrid representation formats can help produce more accurate similarity scores that reflected to the linguistic characteristics of each address element. However, Only through such comprehensive evaluation can one draw reliable conclusions regarding the optimal combination of algorithms and representation formats for next-generation geocoding systems.

The \textbf{geocoding interpolation} component plays a critical role in determining the final geocoded output, as it transforms the matched reference record into geographic coordinates through spatial reasoning. Consequently, the key challenge for this component is to meet the requirements for geocoding outputs, namely, to transparently report the spatial reasoning logic used during interpolation and to quantify the spatial uncertainty associated with the derived location. In this end, a research question can be:
\begin{itemize}
    \vspace{0.25em}
    \item \textit{Q:} How can interpolation methods be designed to enhance their spatial reasoning processes and quantify the spatial uncertainty of the resulting geocode? \\
      \textit{H:} Interpolation approaches must integrate more spatial reasoning capability to handle complex spatial expressions, and they must explicitly model proximity to nearby features in order to generate meaningful uncertainty estimates.
    \vspace{0.25em}
\end{itemize}
We argue that an effective interpolation algorithm must incorporate more advanced spatial reasoning capabilities to handle the complexity of geocoding scenarios such as areas with unknown building distribution patterns or location descriptions containing topological relationships. Recent work by ~\citep{al2019towardsGeocodingSpatialExpressions} made some progress in this direction by addressing simple distance-based expressions relative to points of interest. Another important aspect that new interpolation methods should have is to trace and document their spatial reasoning steps so that the interpolation procedure can be transparently reported and the uncertainty of the final output can be appropriately estimated.

In this part, we present a set of research questions and hypotheses for each geocoding component. We envision that this list will help motivate future experimental investigations and guide the development of next-generation geocoding techniques.

\section{Pathways Toward Next-Generation Geocoding Development}
\label{sec:Opportunities in Geocoding}

Building on the methodologies, challenges, and remaining research questions discussed above, we now explore broader opportunities for advancing next-generation geocoding systems. In contrast to the component-level research directions outlined earlier, the opportunities presented here highlight new pathways or new paradigm for future geocoding system development.

\subsection{Uniformed benchmark framework for geocoding}

As discussed in the previous sections, developing new geocoding techniques or identifying optimal solutions requires systematic performance evaluation. However, a major issue persists: existing evaluation results are often incomparable due to the absence of standardized datasets and metrics \citep{jacquez2012research}. By contrast, many NLP tasks, such as NER, benefit from well-established and standardized annotated datasets \citep{conll2003NERdataset}. To date, existing geocoding evaluation datasets fail to fully reflect real-world input quality. Although geocoding inputs often include both syntactic and semantic errors,  existing evaluations predominantly use simplistic misspellings and overlook more complex errors, such as semantic inaccuracies, leaving system performance uncertain when handling erroneous inputs. Recently, \cite{yin2023chatgpt} introduced a benchmark U.S. address dataset containing over 21 error categories observed in production geocoding services. However, further efforts are still needed to expand this dataset to encompass a wider range of error types, languages, and format variations. 

To benefit future development of geocoding systems, we argue that the foremost priority is to create a unified benchmark framework for geocoding. This framework should have the following features: (1) evaluation data should better represent real-world input quality to more accurately assess geocoding performance; (2) datasets, metrics, and evaluation outputs should be standardized for both component-level and end-to-end geocoding workflows; and
(3) the framework should be made openly accessible to permit future researchers to evaluate individual geocoding sub-workflow or full geocoding systems as needed.

\subsection{A continuously self-improving geocoding system}

It is no secret that web search engines can improve the output by (1) continuously crawling web content to maintain the freshness of their reference datasets and provide the latest search results \citep{lewandowski2008three}, and (2) mining system logs such as input logs, feedback logs to find common input patterns and improve search and recommendation algorithms ~\citep{pan2015feedbackUpateRecSystem}. 
We propose that the next-generation geocoding system should adhere to the similar principle to continuously update its reference dataset, enhance its retrieval capabilities, and re-train its machine learning-based modules to ensure high-quality geocoding outputs. We named this geocoding system as a \textit{self-improving} system. This is crucial for the following reasons. First, rapid urban development, such as new construction buildings, often leads to temporal inaccuracies in reference datasets, making them incomplete or outdated. Second, variations in the descriptions or abbreviations of place names, influenced by time and human cognition, can hinder search accuracy; using aliases for these place names has been shown to improve search outcomes, but manually collecting and updating aliases is inefficient. Third, machine learning modules within geocoding systems may experience performance degradation over time, a phenomenon known as model shifting \citep{chen2021modelshifting}. For example, new input errors and variation patterns may deviate from those in the data used to train a machine learning-based address parser, leading to degraded model performance. Several works have made some progress to mitigate these issues. For example, ~\cite{yin2019DLRooftopGeocoding} applied object detection techniques to geocoding, showing that it can help generate or update address reference data for an input as long as reference records are found for nearby parcels or buildings.
 \cite{yin2023chatgpt} developed a framework to mine input errors and variation patterns from geocoding system logs and generate low-quality geocoding inputs from these patterns. ~\cite{clemens2018indexingSpellErrorforGeocoding} found that the frequency of erroneous input has a Pareto distribution from logs, and re-index the most frequent spelling variant into reference datasets to improve both the recall and the precision of query results. We argue that by leveraging historical geocoding transactions to automatically re-train an address parser and a classifier or to augment the reference dataset, it is possible to achieve a self-improving geocoding system. 
 
However, to fully deliver such a geocoding system, several questions must be addressed in advance. First, when mining geocoding logs, sensitive information, such as contact details or personally identifiable attributes, must be carefully removed to prevent potential privacy leakage. Therefore, a mechanism to perform for K-anonymity or differential privacy needs to be embedded into the workflow so that sensitive information is neither used during model training nor re-indexed into the reference dataset. Second, we need to have a methodology about how to update each geocoding component in a systematical manner to improve overall outcome. For example, we need to have clear answers for the following questions: (1) to what extent can retraining the address parser enhance parsing outcomes, and (2) are there more cost-effective alternatives, such as reorganizing the reference dataset that could achieve similar improvements? Third, once the update strategies are defined, the next challenge is determining when these updates should occur. Given the substantial computational and temporal costs of model training, it is critical to develop a performance evaluation protocol that triggers the self-improvement process only when necessary, thereby ensuring the improvement of the whole geocoding system is achieved at minimal cost.

\subsection{LLMs for an end-to-end geocoding workflow}
 
In previous section, we observe that LLMs have been applied to develop address parsers, retrieval classifiers, and agents capable of performing geocoding tasks. Moving beyond these improvements to individual components, we envision that LLMs will increasingly support geocoding tasks in a more end-to-end manner.
Specifically, for the geocoding textual retrieval phase, LLMs offer an opportunity to shift from traditional address element-based matching to an embedding-driven retrieval framework. Instead of segmenting the input description into indexed address elements and searching the reference dataset accordingly, LLMs can generate a vector representation of the entire address description and retrieve the most similar record directly from the reference dataset. Furthermore, textual similarity and spatial distance can be jointly incorporated into a loss function to learn embeddings that align both linguistic and spatial proximity
. Under such a framework, when a new address description is received, the system would encode the full description into an embedding and then identify the most similar address record based on these learned representations. For the geocoding interpolation workflow, we expect that LLMs can help translate topological relationships into specific spatial calculations, especially, using chain-of-thought reasoning \citep{wei2022chain} to derive a set of step-by-step spatial reasoning procedures. To this end, we can have an end-to-end geocoding workflow that can minimize error accumulation across components and improve adaptability to diverse geocoding inputs. Needless to say, pursuing this direction requires substantial effort in model training, validation, and continuous evaluation. For practitioners, it is therefore essential to carefully assess the trade-offs between the computational and operational costs and the potential performance gains of adopting such an approach. However, this remains an interesting direction that warrants further investigation by future researchers.

\section{Conclusions}
\label{Conclusions}

This survey has outlined the fundamental requirements that modern geocoding systems must satisfy and assessed how well current approaches address those needs. By examining the major functional elements of the geocoding workflow, we highlighted the methodological foundations of each component and identified the gaps that remain. We then formulated targeted research questions and hypotheses to guide systematic evaluation and to support the development of more capable geocoding solutions. Building on these analyses, we also discussed broader opportunities and future directions for advancing the field.

Looking ahead, we anticipate that the evolution of next-generation geocoding systems should follow two major trajectories: enhanced understanding of location descriptions and improved spatial reasoning. The first trajectory should involve incorporating semantic representations, neural parsing models, and hybrid similarity measures to robustly interpret diverse and complex address expressions across languages, formats, and domains. A related extension of this trajectory is to learn joint embeddings for address records that reflect both their textual descriptions and their spatial and topological relationships, enabling retrieval models to capture linguistic similarity and geographic proximity in a unified representation space. The second trajectory emphasizes deeper spatial reasoning by integrating geometric context, boundary awareness, and uncertainty estimation to derive multi-step spatial reasoning procedures that more faithfully reflect the distribution of objects. We invite the research community to pursue these trajectories to advance geocoding systems that not only yield more accurate spatial coordinates, but also better understand and interpret real-world spatial complexity.

\begin{appendices}






\end{appendices}

\bibliography{sn-bibliography}

\end{document}